# Exploring terra incognita in the phase diagram of strongly interacting matter — Experiments at FAIR and NICA

P. Senger

Facility for Antiproton and Ion Research, Darmstadt, Germany

E-mail: p.senger@gsi.de

**Abstract**

The fundamental properties of dense nuclear matter, as it exists in the core of massive stellar objects, are still largely unknown. The investigation of the high-density equation of state (EOS), which determines mass and radii of neutron stars and the dynamics of neutron star mergers, is in the focus of astronomical observations and of laboratory experiments with heavy-ion collisions. Moreover, the microscopic degrees-of-freedom of strongly interacting matter at high baryon densities are also unknown. While Quantum-Chromo-Dynamics (QCD) calculations on the lattice find a smooth chiral crossover between hadronic matter and the quark-gluon plasma for high temperatures at zero baryon chemical potential, effective models predict a $1^{st}$ order chiral transition with a critical endpoint for matter at large baryon chemical potentials. Up to date, experimental data both on the high-density EOS and on a possible phase transition in dense baryonic matter are very scarce. In order to explore this terra incognita, dedicated experimental programs are planned at future heavy-ion research centres: the CBM experiment at FAIR, and the MPD and BM@N experiments at NICA. The research programs and the layout of these experiments will be presented. The future results of these laboratory experiments will complement astronomical observations concerning the EOS, and, in addition, will shed light on the microscopic degrees of freedom of QCD matter at neutron star core densities.

## 1. Introduction

The investigation of strongly interacting matter at high temperatures, as it prevailed in the early universe, or at high baryon densities, as it still exists in compact stellar objects, is in the focus of the research programs with heavy-ion beams at accelerator centres all over the world. At extremely high collision energies, as available at the Relativistic Heavy Ion Collider (RHIC) in Brookhaven, or at the Large Heavy-ion Collider (LHC) at CERN, the experimental observables indicate that an elementary form of matter and antimatter is created in the hot reaction volume, consisting of quarks and gluons, which finally coalesce to hadrons when the fireball expands and cools down. The measured hadron yields can be reproduced by statistical hadronization models by assuming a chemical freeze-out temperature, a baryon chemical potential, and a volume [1]. For heavy-ion collisions at LHC, where particles and antiparticles are produced in equal amounts, the model fit to the data results in a baryon chemical potential of $\mu_B \approx 0$ and a freeze-out temperature between 150 and 160 MeV [1]. For very same values of T and $\mu_B$, Lattice Quantum-Chromo-Dynamics (LQCD) calculations find a smooth crossover from quark-gluon degrees-of-freedom to hadronic matter, caused by the spontaneous breaking of the chiral symmetry of QCD [2,3]. This coincidence suggests, that the chemical freeze-out of hadrons and the hadronization process happen simultaneously, at least for small baryon chemical potentials.

At lower collision energies, where less antiparticles are produced, and, hence, the baryon-chemical potential increases, the freeze-out temperature decreases. Such freeze-out points in the T versus $\mu_B$ diagram have been extracted from measured particle yields by various statistical models [1,4,5]. Whether the freeze-out curve coincides with the pseudo-critical temperature of a crossover phase transition also for finite baryon-chemical potentials, and if yes, up to which value of $\mu_B$, is still an open question. Moreover, it is unclear whether the crossover transition continues to very large $\mu_B$, or ends in a second-order critical endpoint of a first-order transition between quark-gluon plasma and hadron matter [6].

Due to the sign problem in finite density LQCD, this theory is not applicable for the calculation of phase transitions and the search for a critical endpoint at larger baryon-chemical potentials. Recently, LQCD calculations determined an upper bound for the temperature of a critical endpoint at $\mu_B = 0$ in the chiral limit, i.e. at vanishing light quark masses [7,8]. The result is illustrated in figure 1, which depicts a 3-dimensional phase diagram as function of temperature T, baryon-chemical potential $\mu_B$ and mass of the light quarks. The red dot corresponds to the maximum value of the critical temperature $T_c = 132+3-6$ MeV. The figure indicates, that the critical endpoint $T_{cep}$ of a possible 1st order chiral phase transition for finite $\mu_B$ and for finite quark masses, i.e. under conditions as realized in heavy-ion collisions, might be even much lower.

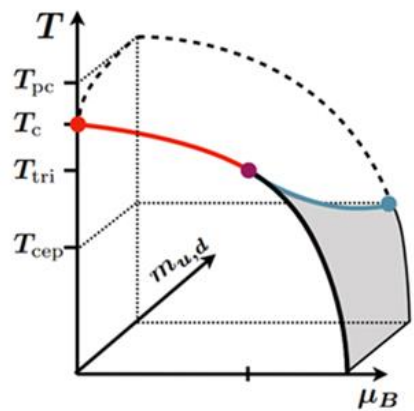

Fig.1: Three-dimensional phase diagram as function of temperature T, baryon-chemical potential $\mu_B$ and mass of the light quarks as calculated by LQCD. Red dot: upper temperature limit $T_c=132+3-6$ MeV for the critical endpoint of a first-order chiral phase transition [7,8]. Taken from [7].

Fireball temperatures of up to 130 MeV will be reached in heavy-ion collisions at energies available at the fixed target mode of the STAR experiment at RHIC, at the future Facility of Antiproton and Ion Research (FAIR) in Darmstadt, and at the Nucleotron-based Ion Collider fAcility (NICA) at the Joint

Institute for Nuclear Research (JINR) in Dubna. In particular, the experiments at FAIR and NICA will perform high precision measurements of observables, which will be studied for the first time at these beam energies, in order to eventually discover signatures of a phase transition in QCD matter at high baryon densities. Some of the observables which are expected to be sensitive to a phase transition and to a critical point will be discussed in this article.

In heavy-ion collisions at FAIR and NICA energies, the baryonic matter in the reaction volume is compressed to at least 5 times saturation density $\rho_0$ [9]. Although these densities are realized only for a very short time within a tiny volume, it is possible to extract information on the high-density equation-of-state (EOS) with the use of microscopic transport model calculations. The EOS is a fundamental property of strongly interacting matter, and the prerequisite for our understanding of neutron stars. In the last two decades, information on the EOS at supra-saturation densities has been extracted from results of heavy-ion collisions experiments performed at beam energies of up to 1.5A GeV at GSI. The relevant observables include the collective flow of protons, light fragments, and neutrons, and yields of kaons measured at subthreshold beam energies [10,11,12,13]. The analysis of the experimental results by microscopic transport models supports a soft EOS up to about 2 $\rho_0$, as it will be shown in the following.

The collective flow of protons has been also measured in Au+Au collisions at beam energies from 2A to 11A GeV at the Alternative Gradient Synchrotron (AGS) at BNL Brookhaven [14]. These energies will be also available at FAIR. According to transport model calculations, which have reproduced the measured proton flow data, only very soft and very hard EOS are excluded up to densities of 5 $\rho_0$ [15].

In the recent years, complementary information on the high-density EOS was also provided by astronomical observations, like the determination of mass and radii of neutron stars [16], and the detection of gravitational waves emitted from neutron star mergers [17]. A recent analysis, combining microscopic nuclear theory calculation, constraints from heavy-ion collision experiments at GSI, and multi-messenger astronomical observations, provided new constraints on the EOS of nuclear matter at densities up to about 2 $\rho_0$ [18]. Future heavy-ion experiment at FAIR and NICA will provide new precision data, which will further constrain the EOS at neutron star core densities. Possible observables will be discussed below.

At high baryon densities, the EOS is expected to be modified by the appearance of hyperons, depending on their $\Lambda N$, $\Lambda NN$ and $\Lambda\Lambda N$ interactions, which are largely unknown [19]. Moreover, model calculations suggest that at baryon densities above about 4 - 5 times saturation density, which prevail in the core of neutron stars, nucleons start to percolate and dissolve into their elementary constituents [20]. Whether such a transition modifies the EOS, depends on details of the interaction between the elementary particles [20,21]. It will be very difficult, to extract information on the role of hyperons and on the microscopic-degrees-of freedom in dense QCD matter from astrophysical observations. In contrast, terrestrial experiments with heavy-ions are able to measure observables, which are sensitive both to hyperon-nucleon interactions, and to phase transitions at high baryon densities, as it will be outlined in the following sections.
.

2. **Unravelling the QCD phase diagram at high baryon densities**

Our present knowledge on the phase diagram of strongly interacting matter is illustrated in figure 2, which depicts the temperature T versus baryon chemical potential $\mu_B$ [22]. The various coloured symbols represent the freeze-out conditions as extracted by statistical hadronization models from particle yields measured at different energies. For LHC collision energies, a freeze-out temperature of T = 156.5 ± 1.5 MeV and a baryon chemical of $\mu_B$ = 0.7 ± 3.8 MeV were determined from the data [1]. For similar values of T and $\mu_B$, LQCD calculations found a smooth chiral crossover from the Quark-Gluon Plasma (QGP) to hadronic matter, and obtained a pseudocritical temperature. The HotQCD collaboration determined a pseudo-critical temperature of $T_c$ = 156.5 ± 1.5 MeV at vanishing $\mu_B$, and extrapolated the crossover region up to $\mu_B \approx$ 300 MeV, as indicated by the orange band in figure 2 [2]. The WB collaboration found a pseudo-critical temperature of $T_c$ = 158.0 ± 0.6 MeV with a width of $\Delta T$=15±1 MeV at $\mu_B$=0, and extrapolated the crossover region from imaginary to real chemical potentials up to $\mu_B \approx$ 400 MeV, as illustrated in figure 2 by the green band [3]. Figure 2 also presents results of Dyson-Schwinger Equations (DSE) and Functional Renormalization Group (FRG), which agree to the LQCD results at $\mu_B = 0$, and predict the location of a 1$^{st}$ order phase transition for larger values of $\mu_B$ together with a critical endpoint [22]. The most recent DSE-FRG calculation for 2+1 flavor QCD represented by the blue-dashed line ends up in a critical endpoint at a temperature of $T_{cep}$=93 MeV and a baryon-chemical potential of $\mu_{Bcep}$= 672 MeV. Such conditions are realized in the reaction zone of heavy-ion collisions at a beam kinetic energy between 4A and 6A GeV, which are available in the fixed target mode at the STAR experiment at RHIC, and will be available at the FAIR facility.

It is worthwhile to note, that the DSE-FRG calculations are corroborated by the results of LQCD calculations presented in the introduction, which determine an upper bound for the critical temperature of a possible chiral transition to about $T_{cep}$ = 130 MeV at $\mu_B$=0. Under realistic conditions, i.e. at finite values of $\mu_B$ and for non-zero quark masses, the temperature of the critical endpoint should be even lower.

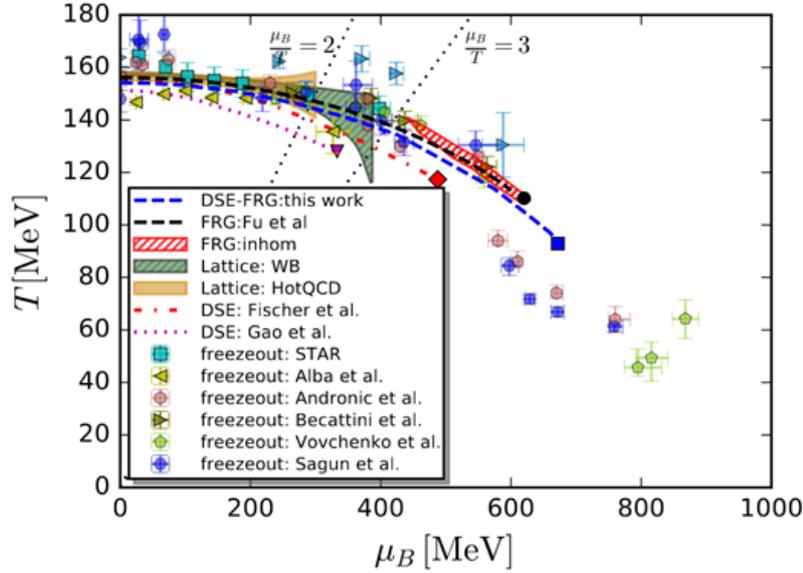

Fig.2: Two-dimensional phase diagram for $N_f$ = 2+1 flavor QCD in comparison to other theoretical approaches and phenomenological freeze-out data. For details see [22]. The DSE-FRG calculation represented by the blue-dashed line predicts a critical endpoint of a 1$^{st}$ order chiral phase transition at $T_{cep}$ = 93 MeV and $\mu_{Bcep}$ = 672 MeV.

In the following, some observables will be discussed, which are expected to be sensitive to a change of microscopic properties of dense QCD matter produced in heavy-ion collisions.

2.1. Chemical equilibration of multi-strange hyperons

As discussed above, the chemical freeze-out temperature extracted by statistical hadronization models from particle yields measured at LHC coincides with the pseudo-critical temperature of a chiral crossover calculated by LQCD. In particular, the chemical equilibration of multi-strange (anti-) hadrons is regarded as a direct consequence of a crossover transition, as it requires very high densities, which are expected to occur close to the phase change [23]. It was argued, that in the short-lived hadronic phase well below the pseudocritical temperature, the hyperon-nucleon cross sections are too small in order to drive multi-strange (anti-) hadrons into chemical equilibrium. Also in heavy-ion collisions at much lower beam energies such as 30A GeV, the Ξ and anti-Ξ hyperons are found to be chemically equilibrated at a freeze-out temperature of T = 138 MeV and a baryon chemical potential of $\mu_B$ = 380 MeV [24]. It is worthwhile to note, that the matter in the reaction volume of a Pb+Pb collision at this moderate beam energy also undergoes a chiral crossover, according to the LQCD [7] and DSE-FRG [22] calculations discussed above. Ξ hyperons have been even measured in Ar + KCl collisions at an energy as low as 1.76A GeV [25], which is almost 2 GeV below the threshold for Ξ production in nucleon-nucleon collisions. In this case, a chemical freeze-out temperature of 70 ± 3 MeV and a baryon chemical potential of 748±8 MeV have been extracted from the yields of produced particles. The Ξ⁻ hyperon yield, however, exceeds the prediction of the statistical hadronization model by a factor of 24±9 [25]. At the future accelerator centres FAIR and NICA, the excitation functions of $\Xi^\pm$ and $\Omega^\pm$ hyperons will be carefully studied in heavy-ion collisions at beam energies between 2 A and 30A GeV, in order to search for the energy, where the multi-strange hyperons are driven into chemical equilibration.

2.2. Probing quark degrees-of-freedom at high baryon densities with charmonium

The suppression of J/ψ mesons in heavy-ion collisions was one of the first proposals for an unambiguous evidence of the formation of the Quark-Gluon Plasma [26]. The underlying effect was color screening of the c c-bar pair in a deconfined medium. Many years later, the charmonium yield was measured by the NA50 collaboration at SPS in Pb+Pb collisions at 158A GeV [27]. When plotting the observed J/ψ yield as function of the energy density ε, the yield declined steeply above a value of ε ≈ 2.5 GeV/fm$^3$. This effect was interpreted as confirmation of the proposed J/ψ suppression in the QGP. An alternative scenario, but also confirming the existence of the QGP at SPS energies, was proposed by the Statistical Hadronization Model (SHM): the charm and anti-charm quarks are produced in hard scattering processes, then thermalize in the QGP, and finally form J/ψ mesons by statistical hadronization at the phase boundary [28]. Within the SHM also prediction for the yield of J/ψ meson in heavy-ion collisions at LHC energies have been made. Due to the much higher energy, the J/ψ yield is expected to be enhanced in heavy-ion collisions compared to a superposition of independent nucleon-nucleon collisions [28,29]. Finally, also in Pb+Pb collisions at LHC energies a slight suppression of J/ψ mesons was found by the ALICE experiment, less pronounced than at SPS or RHIC energies [30].

An important question is, whether charmonium could also serve as a diagnostic probe of a possible phase transition in heavy-ion collisions at lower beam energies and at larger baryon-chemical potentials. According to SHM calculations, the ratio of (J/ψ)/(D+anti-D) is almost independent of beam energy, because the charm quark production in hard collisions is decoupled from the hadronization process resulting in charmed mesons [29]. In hadronic matter, however, the production thresholds for J/ψ, D-bar and D mesons in nucleon-nucleon collisions are $E_{thr}$ = 11.3 GeV, 11.9 GeV, and 14.9 GeV, respectively. Therefore, in heavy-ion collisions close to these thresholds, the (J/ψ)/(D+D-bar) ratio increases strongly towards lower beam energies, as demonstrated by calculations with the Hadron-String-Dynamics (HSD) code [31]. Up to now, J/ψ mesons have not been observed in heavy-ion collisions below top SPS energies. The (J/ψ)/(D+D-bar) ratio will be measured at the NICA collider in Au+Au collisions up to $\sqrt{s_{NN}}$ = 11 GeV. At FAIR, Au beams can be accelerated up to the J/ψ production threshold, but higher energies are available for beams of lighter nuclei.

In a hadronic scenario, charmed particles might also be produced in heavy-ion collisions at subthreshold beam energies. Within the UrQMD event generator, hadrons with charm quarks can be created via the decay of nucleonic resonances, which are excited in multiple hadronic collisions [32]. The resulting yields of J/ψ mesons and $\Lambda_C$ hyperons produced in central Au+Au collisions are depicted in figure 3 as function of collision energy, together with the yields for p+p collisions. In addition, results of HSD calculations are shown in figure 3. The UrQMD calculations demonstrate that in Au+Au collisions at FAIR, total J/ψ multiplicities in the order of 10$^{-6}$ can be reached in a central Au+Au collision, which would be sufficient for a charmonium measurement. A corresponding feasibility study is discussed in chapter 4. It is important to note, that the identification of J/ψ mesons in heavy-ion collisions at FAIR energies would rule the existence of a phase transition in the respective region of the QCD phase diagram, which would be a major discovery, in view of the theoretical predictions on the location of the critical endpoint discussed above.

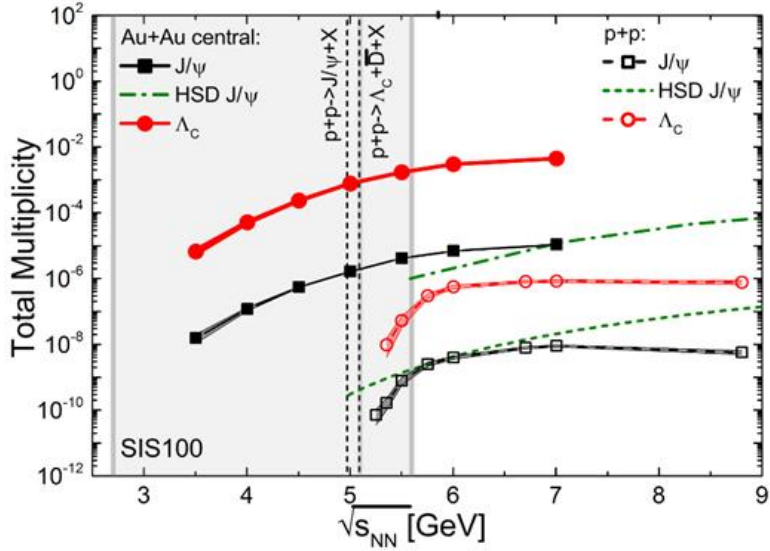

Fig. 3: Yields of J/ψ mesons and $\Lambda_C$ hyperons in p+p and central Au+Au reactions as a function of the collision energy. Green lines: Results of HSD calculations. Vertical lines: Threshold energies for p+p collisions. Grey Area: SIS100 energy. Taken from [32].

2.3. Searching for a phase transition with dileptons

Lepton pairs are considered to be a very sensitive probe of the properties of hot and dense matter produced in heavy-ion collisions, but their measurement is an experimental challenge because of the huge combinatorial background of uncorrelated leptons. Dileptons produced in heavy-ion collisions leave the fireball unaffected by strong interaction, and, hence, carry direct information on the hot and dense matter they are created in. Dilepton sources of particular interest are vector mesons and thermal radiation. The invariant mass spectrum of $e^+e^-$ or $\mu^+\mu^-$ pairs below about 1 GeV/$c^2$ is populated by pion Dalitz decays, and by pairs from ω, ρ, and φ vector meson decays. Lepton pairs from the short lived ρ meson, which decays inside the fireball, permit to extract the in-medium mass distribution of the ρ meson, which is broadened due to the restoration of chiral symmetry in dense matter.

The invariant mass spectrum of $e^+e^-$ or $\mu^+\mu^-$ pairs above about 1 GeV/$c^2$ is not more contaminated by vector meson decays, and, hence, provides the possibility to study the average temperature of the emitting source. This temperature can be directly extracted from the inverse slope of the dilepton invariant mass spectrum [33]. The unique feature of this method is the fact that the slope of the dilepton invariant mass spectrum is not blue-shifted by the radial flow of the emitting source, as it is the case for the slope of hadron momentum and energy distributions. The NA60 collaboration has measured the $\mu^+\mu^-$ invariant mass spectrum in In+In collisions at 158A GeV, and extracted a temperature of T = 205 ± 12 MeV from the slope above 1 GeV/$c^2$ [34]. This value represents an average temperature integrated over the fireball evolution. In principle, it is also possible to extract the average temperature of the fireball from the dilepton invariant mass spectrum at masses below 1 GeV/$c^2$, if lepton pairs from known sources are carefully subtracted. This procedure has been performed by the HADES collaboration, which measured $e^+e^-$ pairs in Au+Au collisions at a beam kinetic energy of 1.25A GeV, and extracted an averaged fireball temperature of T = 72 ±2 MeV [35].

The measurement of the average temperature of the hot and dense fireball as function of beam energy opens the unique opportunity to determine the caloric curve of QCD matter, if it features a 1$^{st}$ order phase transition at all. As illustrated in figure 2 and discussed above, such a possible chiral transition should be located at temperatures and baryon chemical potentials, which can be reached in heavy-ion collisions between 2A and about 10A GeV. In this beam energy range, no dilepton measurements have been performed yet in heavy-ion collisions. Therefore, future dilepton experiments at FAIR and NICA will play an unrivalled and decisive role concerning the discovery or exclusion of the long sought 1$^{st}$ order phase transition in hot and dense QCD matter.

2.4. Fluctuations of proton number distributions

Event-by-event fluctuations of conserved quantities such as net baryon number, net charge and net strangeness have been proposed as convincing experimental signatures for the critical endpoint of a 1$^{st}$ order phase transition in hot and dense QCD matter [36]. These critical fluctuations increase as the correlation lengths increases in the vicinity of the critical endpoint, a similar phenomenon like critical opalescence in classical binary liquids. The STAR collaboration performed a systematic study of event-by-event fluctuations in central Au+Au collisions within their beam energy scan program, using the net-protons as a proxy for the net baryons [37]. Due to the dynamical evolution and the finite size of the fireball, the STAR collaboration studied higher moments of the proton number distribution, which exhibit a stronger dependence on the correlation length. The result of this study is shown by the red symbols in figure 4, which depicts the ratio of the 4$^{th}$ to the 2$^{nd}$ cumulant $C_4/C_2$ (= kurtosis×variance $\kappa\sigma^2$). The points exhibit a non-monotonic variation with a significance of 3.1$\sigma$ at the lowest collision energy of $\sqrt{s_{NN}}$ = 7.7 GeV.

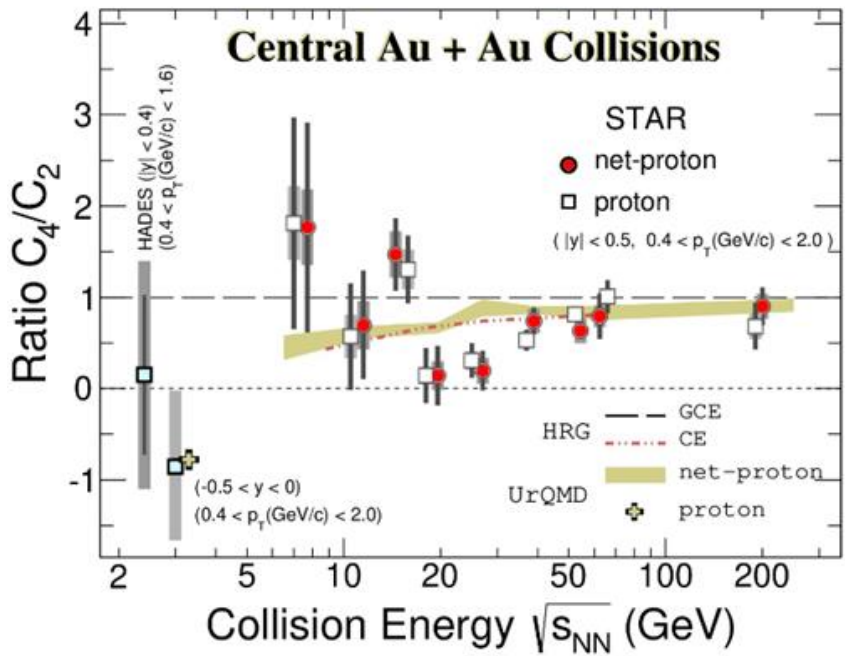

Fig. 4: Ratios of cumulants C4/C2, for proton (squares) and net-proton (red circles) from central Au+Au collisions at RHIC [37]. The plot includes a data point from HADES at $\sqrt{s_{NN}}$ = 2.4 GeV [38], and the recent result of STAR in the fixed target mode at $\sqrt{s_{NN}}$ = 3.0 GeV [39].

The figure also includes the result of the HADES collaboration measured in the 10% most central Au+Au collisions at $\sqrt{s_{NN}}$ = 2.4 GeV, which is compatible with zero [38]. The most recent result of the STAR collaboration obtained in central Au+Au collisions in the fixed target mode at an energy of $\sqrt{s_{NN}}$ = 3.0 GeV, which is compatible with -1, is also shown in figure 4 [39]. In order to finally conclude on the existence of the critical endpoint of a first order phase transition, further high precision measurements of event-by-event fluctuations in the beam energy between HADES and the STAR collider mode are required, which reduce the error bars of the existing data and add new data points. Such measurements will be performed at the future facilities FAIR and NICA. In addition, fluctuations of the yield ratios of light nuclei such as $N_t \cdot N_p/N_d^2$ will be studied, which also have been proposed as signatures for the critical endpoint [40].

3.  **The high-density equation-of-state**

The EOS of nuclear matter can be expressed as pressure $P = \rho^2 \, d(E/A)/d\rho$, where $\rho$ is the density and E/A the binding energy per nucleon. The dependence of the binding energy per nucleon from the density is presented in figure 5 according to various model calculations [13]. The lower set of curves describes the EOS for isospin-symmetric nuclear matter, while the upper set of curves delineates pure neutron matter. The binding energy for isospin symmetric matter, as it exists in atomic nuclei, has a minimum at saturation density $\rho_0$ with E/A = -16 MeV. The curvature of the EOS at $\rho_0$ can be parameterized by the nuclear incompressibility $K_{nm}= 9\rho^2 \, \delta^2(E/A)/\delta\rho^2$, which has been determined for saturation density from the measurement of giant monopole resonances to $K_{nm}(\rho_0) = 230\pm10$ MeV [41].

The EOS at supra-saturation densities, which determines the structure of compact stellar objects and the dynamics of neutron star collisions, is studied since more than three decades in heavy-ion collisions at intermediate beam energies. In particular experiments at the SIS100 accelerator of GSI, where densities of more than 2 $\rho_0$ are reached in the collision zone of heavy-ion reactions, provided constraints on the EOS of symmetric nuclear matter. An important observable is the collective flow of particles, which is generated by the pressure gradient in the reaction volume. The FOPI collaboration at GSI measured the elliptic flow of protons and light fragments in Au+Au collisions at energies from 0.4 to 1.5A GeV [10]. These reactions have been also calculated with the IQMD event generator, which includes momentum dependent interactions and in-medium cross sections. When assuming a nuclear incompressibility of $K_{nm}$=190 ± 30 MeV, IQMD calculations could reproduce the measured flow data.

Complementary to the elliptic flow, subthreshold kaon production in heavy-ion collisions turned out to be an observable sensitive to the EOS of symmetric matter. With this goal, pioneering measurements have been conducted by the KaoS collaboration at GSI. The production of $K^+$ mesons in elementary collisions like p+p → $K^+\Lambda$ p requires a proton energy of 1.58 GeV. In heavy-ion collisions, however, kaons are produced at much lower beam energies due to the Fermi energy and multiple collisions of hadrons, including pions and baryonic resonances. The probability of such collisions increases with increasing softness of the EOS. It is worthwhile to note, that according to simulations, kaons are produced at SIS100 energies predominantly above 2 $\rho_0$. The KaoS collaboration performed systematic measurements of $K^+$ mesons in Au+Au collisions at beam energies from 0.8A GeV to 1.5A GeV, and in C+C collisions from 0.8A GeV to 2A GeV [12]. The investigation of the much smaller collision system serves as an important reference, as the influence of the EOS on kaon production is negligible, while the contributions of medium effects like Fermi motion and of systematic errors on the kaon production are reduced, when comparing the kaon yield ratios $K^+$(Au+Au)/$K^+$(C+C). The same argument holds for the results of microscopic transport model calculation. In this case, also uncertainties in the description of momentum dependent interactions and in-medium cross sections on kaon production are reduced. The RQMD and IQMD transport models extracted a value of $K_{nm} \approx$ 200 MeV from the ratio of the measured

$K^+$ multiplicities [13,42]. The combined results of the FOPI and KaoS measurements on the EOS for isospin-symmetric matter are illustrated in figure 5 by the lower green area.

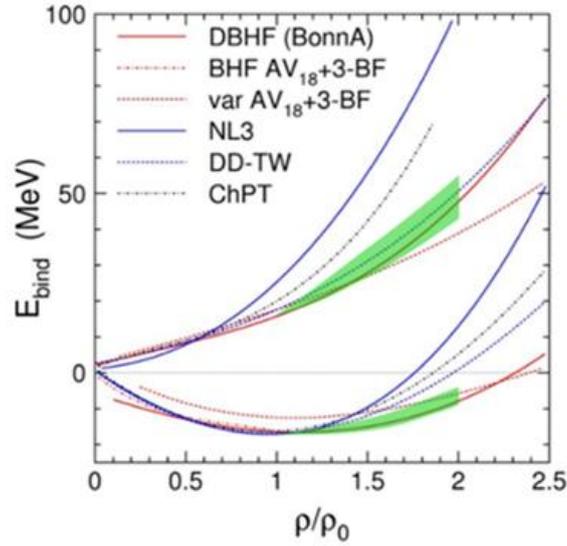

Fig. 5: Binding energy versus baryon density. Lower set of curves: Isospin-symmetric nuclear matter. Upper set of curves: Pure neutron matter [13]. Lower green area: Analysis of experimental data taken at GSI by FOPI and KaoS [10,12,13,42]. Upper green area: Symmetry energy measured by ASY-EOS [43] added to the lower green area (see text).

The EOS for neutron matter and isospin-symmetric matter differs by the symmetry energy $E_{sym}$, which can also be determined in laboratory experiments. In order to contribute to our understanding of neutron stars, $E_{sym}$, should be determined also at supra-saturation densities, i.e. in heavy-ion collisions. In this case the relevant observables, which exhibit a sensitivity to $E_{sym}$, are the flow of neutrons, and particles with opposite third component of the isospin $I_3$ such as pions [44].The elliptic flow of neutrons and protons has been measured by the ASY-EOS collaboration at GSI in Au+Au collisions at a beam energy of 0.4A GeV [43]. The experimental data have been reproduced by UrQMD model calculations, which include the parametrization $E_{sym}(\rho) = E_{sym}(\rho_0) + L/3((\rho - \rho_0)/\rho_0)$. The data are reproduced by assuming a symmetry energy at saturation density of $E_{sym}(\rho_0) = 34$ MeV, and a slope of $L = 72 \pm 13$ MeV for $E_{sym}$ at $\rho_0$ [43]. When adding the ASY-EOS result for $E_{sym}(\rho)$ to the combined result of FOPI and KaoS for the EOS of symmetric matter, one obtains the upper green band in figure 5, which represents the present heavy-ion constraint on the EOS for neutron matter up to densities of 2 $\rho_0$.

Pioneering measurements of the directed and elliptic flow of protons have been performed in Au+Au collisions at energies up to 11A GeV at the AGS in Brookhaven, where baryon densities above 5 $\rho_0$ are reached [14,45,46]. The results have been analyzed with relativistic transport calculations, which extracted a soft EOS ($K_{nm} = 210$ MeV) from the directed flow data, and a stiff EOS ($K_{nm} = 300$ MeV) from the data on elliptic flow [15]. Therefore, only very soft and very stiff EOS are excluded from the flow data measured in heavy-ion collisions so far.

As mentioned in the introduction, constraints on the high-density equation-of-state (EOS) have been also obtained from astronomical measurements of mass and radii of neutron stars [16] including the masses of massive neutron stars [47,48,49], and from the detection of gravitational waves emitted from neutron star mergers [17]. The result of a combined analysis of these astronomical data is presented as orange area in figure 6, which depicts the pressure as function of density for neutron matter. The blue area represents the combined results of the GSI experiments FOPI, KaoS and ASY-EOS, and the pink and green areas correspond to the analysis of the AGS flow data, after adding the symmetry energy with a strong and a weak density dependence, respectively [15]

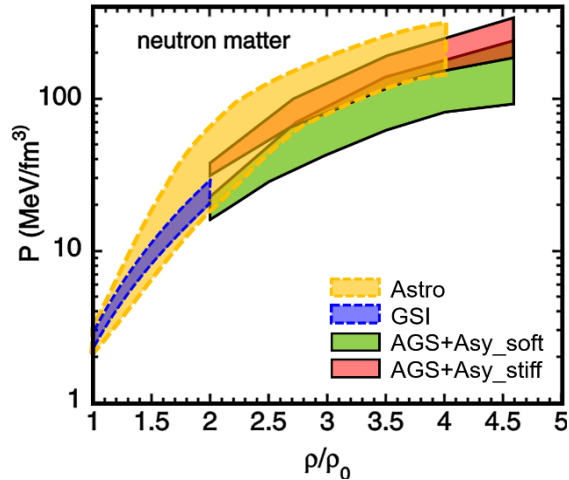

Fig. 6: EOS for cold neutron matter plotted as pressure versus density. Orange area: Analysis of masses and radii of neutron stars measured by NICER, including mass measurements of the most massive neutron stars, and the tidal deformability estimate from the GW170817 [16]. Blue area: Analysis of the GSI flow and kaon data [10,12,13,42,43]. Pink and green area: analysis of AGS flow data [14,45,46] after adding the pressure from asymmetry terms with strong and weak density dependences, respectively.

Figure 6 illustrates, that the EOS up to $2\rho_0$ is constrained by heavy-ion experiments at GSI, while at higher densities the measured masses of the most massive neutron stars exclude a soft EOS as represented by the green area. New measurements of radii and masses of neutron stars together with future laboratory experiments at FAIR and NICA are required in order to further constrain the EOS at neutron star core densities.

Future laboratory experiments will address the high-density EOS both of isospin-symmetric and neutron matter. Detailed measurements of the directed and the elliptic flow of protons and light fragments will be performed. In particular data on the elliptic flow have to be improved: the FOPI data on elliptic flow measured in Au+Au collisions up to beam energies of 1.5A GeV clearly favor a soft EOS, whereas the elliptic flow measured at AGS at 2A GeV is compatible with a hard EOS. For beam energies above 4A GeV, the sensitivity of the elliptic flow on the EOS decreases. In contrast, the sensitivity of the directed flow on the EOS increases with increasing beam energy from 1A to 10A GeV. Therefore, high precision measurements of both the directed and the elliptic flow from 2A to 10A GeV are required, in order to reduce the existing discrepancy between the interpretations of the directed and elliptic flow data, and to futher constrain the EOS for symmetric matter.

Complementary information on the high-density EOS will be provided by subthreshold particle production. As discussed above, this method has been successfully applied with kaons at beam energies up to 1.5A GeV. For higher beam energies from 2A to10A GeV multi-strange (anti-) hyperons are the preferential probes, as the production thresholds in p+p reactions for $\Xi^-$, $\Omega^-$, $\Xi^+$, and $\Omega^+$ hyperons are 3.7, 7.0, 9.0, and 12.7 GeV, respectively. At subthreshold beam energies, the hyperons are produced in heavy-ion collisions via multiple collisions, like the kaons at lower energies. In the production of multi-strange hyperons, also strangeness exchange reactions involving $\Lambda$, $\Sigma$, or $\Xi^-$ hyperons play an important role [50,51]. The sensitivity of the yield of hyperons on the EOS for symmetric matter has been studied with the PHQMD transport model for central Au+Au collisions at a beam energy of 4A GeV [52]. The result of the calculation is presented in figure 7, which depicts the yield ratios of multi-strange (anti-) hyperons for a soft over a hard EOS versus their production threshold energy in p+p collisions. The calculation clearly demonstrates, that more hyperons are produced when the EOS is soft, when higher densities are achieved in the fireball. Moreover, the sensitivity of the EOS increases with increasing mass of the hyperons, i.e. with the "subthresholdness" of the production mechanism. A reference calculation of central C+C collisions at 4A GeV has revealed no dependence of the hyperon yield on the EOS. Up to date, only a few $\Xi^-$ hyperons have been measured in heavy-ion collisions up to SPS beam energies. In conclusion, a systematic study of multi-strange (anti-) hyperon production in heavy and light ion collisions up to beam energies of 10A GeV will further constrain the EOS of symmetric nuclear matter up to densities of 5 $\rho_0$.

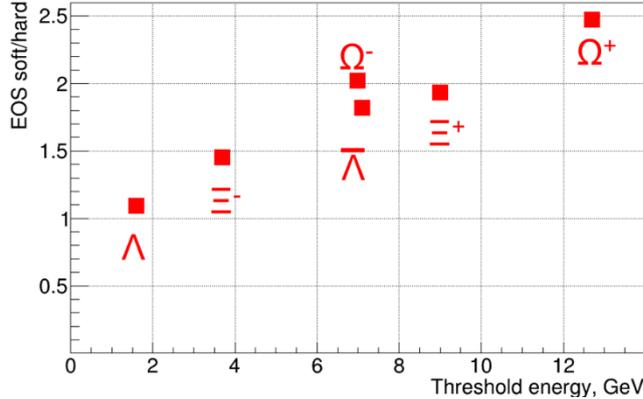

Fig. 7: Hyperon yield ratios for a soft over a hard EOS calculated for central Au+Au collisions at a beam energy of 4A GeV with the PHQMD transport code versus the production threshold in nucleon-nucleon collisions [52].

In order to contribute to our understanding of the EOS for neutron stars, laboratory experiments have to measure also the symmetry energy up densities of about 5 $\rho_0$. As discussed above, the elliptic flow ratio of neutrons over protons has been proven to be sensitive to $E_{sym}$ up to densities of about 2 $\rho_0$. Information on $E_{sym}$ at densities slightly above saturation density has been obtained from the pion ratio $\pi^-/\pi^+$ measured as function of transverse momentum in the neutron rich collision system $^{132}$Sn + $^{124}$Sn [53]. Promising probes of $E_{sym}$ at baryon densities above 2 $\rho_0$ might be $\Sigma^-$(dds) and $\Sigma^+$(uus) hyperons, which differ in the third component of the isospin $I_3 = \pm 1$, and reflect the density of neutrons (ddu) and protons (uud). The identification of $\Sigma^\pm$ hyperons is experimentally challenging, as one of the decay daughters is a neutral particle. Alternatively, the ratio of excited $\Sigma^{*-}$(dds) and $\Sigma^{*+}$(uus) hyperons can be studied, which can be easily identified via their decay into $\Lambda\pi$ pairs.

As illustrated in figure 6 by the blue area, precise laboratory measurements with heavy-ion collisions at GSI can substantially improve the constraints on the EOS, and complement the analysis of astronomical observations. Future experiments with heavy-ions at intermediate energies will also further constrain the EOS at neutron star core densities. In contrast to astronomical observations, heavy-ion experiments allow to address even more fundamental aspects concerning the microscopic degrees-of-freedom and their interactions in dense matter, which finally determine the EOS. For example, calculations based on the concept of hadron-quark continuity, where hadronic matter transforms smoothly into quark matter with increasing density, provide a EOS for neutron stars from the solid crust through the liquid nuclear matter regime to the quark core. At high densities, the repulsive interaction between the quarks result in a stiff EOS, which is required to stabilize massive neutron stars [54]. Such a cross over transition from hadronic to quark matter can be studied in heavy-ion collisions, as discussed in the previous section. Moreover, the role of hyperons, which are expected to appear in dense neutron star matter if the chemical potential of neutrons exceeds the chemical potential of hyperons, is not yet resolved. The resulting softening of the EOS would prevent the existence of massive neutron stars, which have been observed. This "hyperon puzzle" can be resolved by assuming strongly repulsive nucleon-hyperon interactions together with three-body forces involving a hyperon and two nucleons, which shift the condensation of hyperons up to densities beyond 5 $\rho_0$ [19,55]. Experimental information on the $\Lambda$N and $\Lambda$NN interaction can be obtained by the measurement of masses and lifetimes of light lambda hypernuclei, which are abundantly produced in heavy-ion collisions at FAIR energies [56]. The study of light double-lambda hypernuclei, which are also created in such collisions although with much lower cross-sections, will shed light on three-body interactions involving two hyperons.

## 4. Future experiments at FAIR and NICA

Several experimental setups are being built at future heavy-ion accelerator centres in order to explore the QCD phase diagram in the region of large baryon chemical potentials, and to constrain the EOS at neutron star core densities. At FAIR in Darmstadt, Germany, the Compressed Baryonic Matter (CBM) experiment is under construction, which is designed to perform high precision experiments of bulk observables and rare probes at extremely high reaction rates in the energy range up to 11A GeV for Au-beams. At NICA at JINR in Dubna, Russia, the Multi-Purpose Detector (MPD) will measure hadrons and electron pairs in heavy-ion collisions up to collisions energies of $\sqrt{s_{NN}}$ = 11 GeV (≈ 63A GeV). In addition, the Baryonic Matter experiment at the Nuclotron (BM@N) at JINR is being upgraded in order be able to measure hadrons in Au+Au collisions up to 3.8A GeV. The reaction rates as function of collision energy for these experiments are depicted in figure 8, together with the capabilities of existing detector setups like HADES at GSI, NA61/SHINE at the CERN SPS, and STAR at RHIC, both for the collider and the fixed target mode. The collider experiments STAR and MPD suffer from decreasing luminosity for energies below the design values. The experiments at FAIR and NICA complement each other in a collision energy range, where the highest baryon density is created in heavy-ion collisions. In the following, the layout of these experiments will be presented.

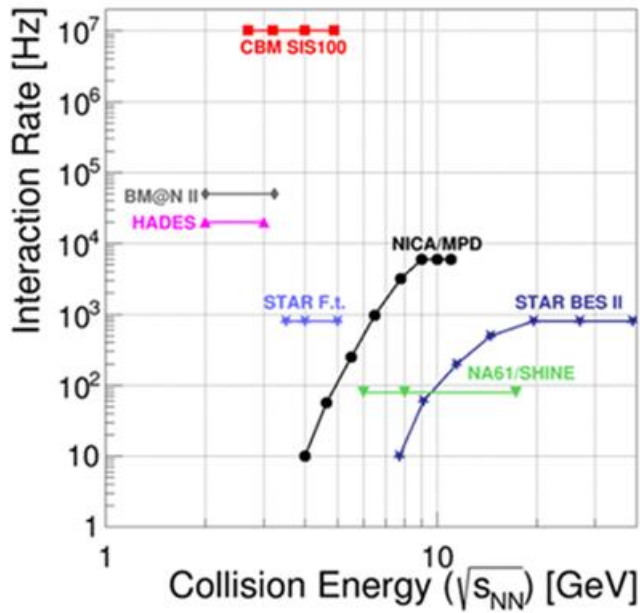

Fig. 8: Rates of Au+Au reactions as function of collision energy for existing experiments and experiments under construction (see text, taken from [57]).

4.1 The Facility for Antiproton and Ion Research

The future Facility for Antiproton and Ion Research (FAIR) covers a broad range of fundamental and applied research programs based on particle beams [58]. The workhorse of FAIR is the SIS100 synchrotron (magnetic rigidity 100 Tm), which accelerates for example $^{238}U^{92+}$ ions up to energies of 11A GeV with an intensity of up to $10^{10}$ ions/s, and protons up to 29 GeV with an intensity of up to $3 \cdot 10^{13}$/s. Alternatively, beams of $^{238}U^{28+}$ ions can be provided with energies of 1.5A GeV and intensities up to $10^{12}$ ions/s. These beams will be used for the production of short-lived rare isotopes, which will be selected in-flight by a Superconducting Fragment Separator (SFRS), and transported to various detector systems, where experiments on nuclear structure and nuclear astrophysics take place. The high-intensity proton beam will produce beams of antiprotons with momenta between 1.5 and 15 GeV/c and intensities up to $10^{11}$/s, which will be used for hadron physics experiments with the PANDA detector.

A variety of experiments on atomic physics, plasma physics and applied research in biophysics, medical science, and material sciences will be performed by the APPA collaboration. The fourth scientific pillar of FAIR is dense nuclear matter physics, which will be conducted by the HADES and the CBM experiments, which will be discussed below. A sketch of GSI and FAIR is presented in figure 9, and the progress in FAIR civil construction is illustrated in figure 10. First FAIR experiments are planned for 2025.

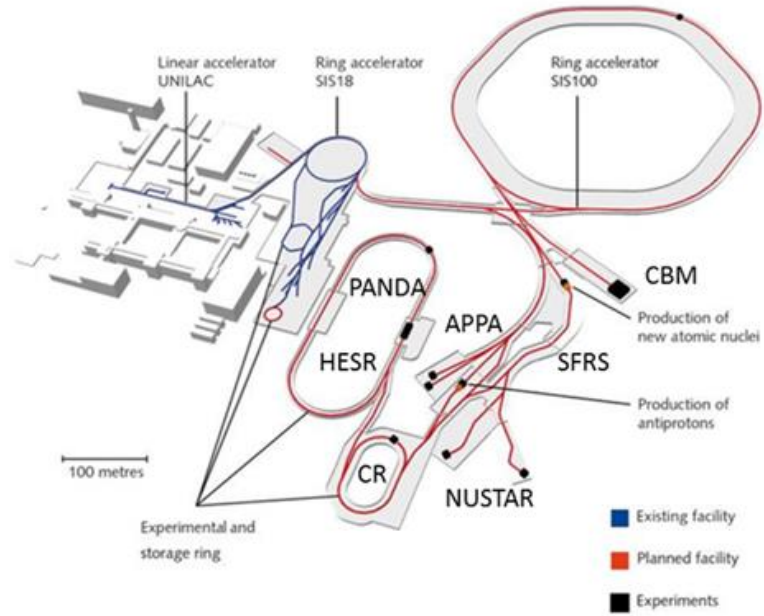

Fig. 9: Sketch of GSI and FAIR with the Synchrotrons SIS18 and SIS100, the High-Energy Storage Ring (HESR), the Collector Ring (CR), the Super-conducting Fragment Separator (SFRS), and the experimental sites of NUSTAR (**Nu**clear **St**ructure, **A**strophysics and **R**eactions), PANDA (Anti**P**roton **An**nihilation at **Da**rmstadt), APPA (**A**tomic **P**hysics, **P**lasma physics and **A**pplied research), and CBM (**C**ompressed **B**aryonic **M**atter).

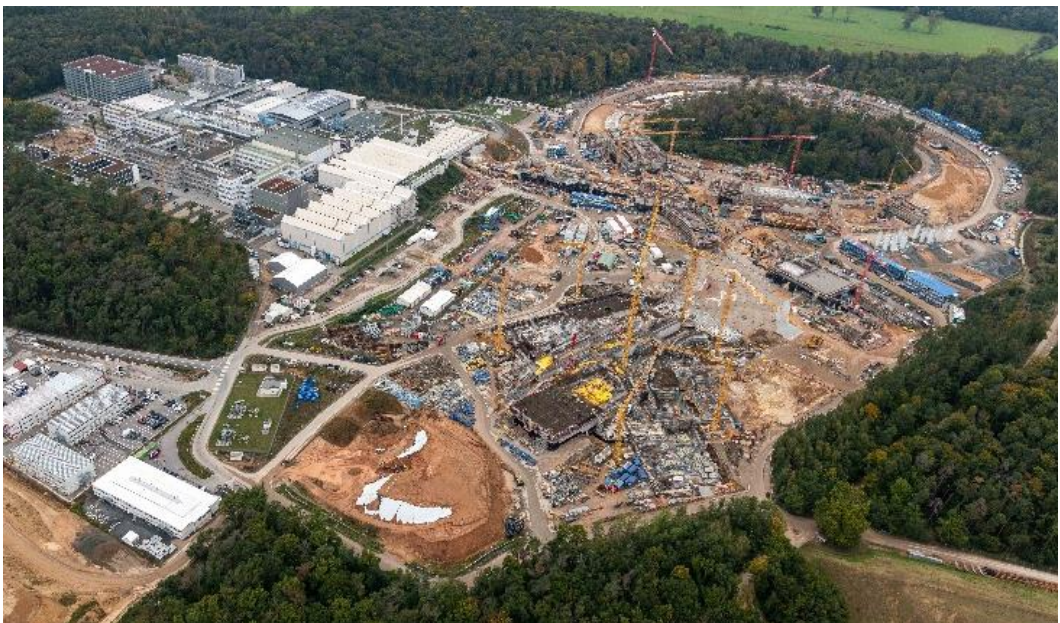

Fig. 10: Photo of the FAIR construction site from October 2021 [59].

## 4.2 The CBM experiment at FAIR

In order to contribute to the solution of the open questions discussed above, the CBM detector system is designed to measure light and heavy collision systems at beam energies from 2A to 11A GeV, and to perform precision studies of

- multi-differential hadron distributions

- event-by-event fluctuations of hadron multiplicities

- correlations between hadrons

- collective flow of identified particles,

- multi-strange (anti-) hyperons

- lepton pairs ($e^+e^-$ and $\mu^+\mu^-$)

As illustrated in figure 8, the CBM experiment will be operated at unprecedented heavy-ion collision rates of up to 10 MHz. Such a performance can only be achieved with fast and radiation hard detectors. Charged particle tracking will be performed by a Micro-Vertex Detector (MVD) and a Silicon Tracking System (STS), both positioned in the large gap of a superconducting dipole magnet. The MVD consists of four stations with Monolithic Active Pixel Silicon Sensors (MAPS), located up to 20 cm downstream the target. The eight stations of the STS are equipped with 900 double-sided micro-strip silicon sensors, positioned from 25 cm and to 100 cm downstream the target. The identification of electrons and positrons is performed by two detector systems: the Ring Imaging Cherenkov (RICH) detector located behind the STS and the magnet, and a Transition Radiation Detector (TRD) further downstream. The electron identification efficiency of the RICH decreases above electron momenta above 10 GeV/c, whereas the TRD is more efficient above this value. For muon measurements, the RICH will be substituted by a Muon Chamber (MuCh) system, while the TRD stays in place and provides additional tracking of the muon candidates towards the Time-of-flight detector (TOF). The MuCh consists of a combination of up to five hadron absorbers and four tracking chamber triplets. The first absorber consists of a 60 cm thick block of carbon/concrete, while the following four absorbers are made of iron with thicknesses of 20 cm, 20 cm, 30cm, and 100 cm. Behind the first and second absorber, Gas Electron Multiplier (GEM) detector triplets are installed, while behind the third and fourth absorber Resistive Plate Chamber (RPC) triplets are located. The rate capability per area of the detector systems is adjusted to the particle multiplicity, which decreases with increasing absorber material. The TOF wall is located about 8 m downstream the target, and consists of Multi-Gap Resistive Plate Chambers (MRPC) covering an active area of 120 m$^2$. The polar angle acceptance of the CBM detector setup ranges from about 3° to 25°, and covers midrapidity for the FAIR beam energies. About 10 m downstream the target, the Project Spectator Detector (PSD) is located, which is a hadron calorimeter measuring projectile fragments, which contain information on the orientation of the reaction plane. The CBM detector setup is sketched in the right part of figure 11, while in the left part the High-Acceptance Di-Electron Spectrometer (HADES) is shown. Due to its large polar angle acceptance from 18° to 85°, HADES is well suited for reference measurements such as proton-nucleus collisions and collisions between low-mass nuclei, and will provide data on hadron and di-electron production.

In order to operate the detector system at extremely high reaction rates, the CBM experiment is equipped with a novel data readout and acquisition system. Because of the complicated decay topologies of most of the diagnostic probes, no selective hardware trigger pattern can be generated. Instead, the each single detector signal will get a time stamp from the readout ASIC. The following readout chain then will compress the data, and send it via optical links to the "GeenIT cube", the high-performance computing centre of GSI. Here, high-speed computing algorithms will first reconstruct the charged particle tracks from the position and time information of the signals. In a next step, the mass and momentum of the particles will be determined using information on the time-of-flight and track topology. In the final step, the events will be first reconstructed based on the time correlation of the tracks and their primary vertex, and then selected depending on particular observables. The entire analysis chain has to be performed in real-time, in order to reduce the data volume to be stored.

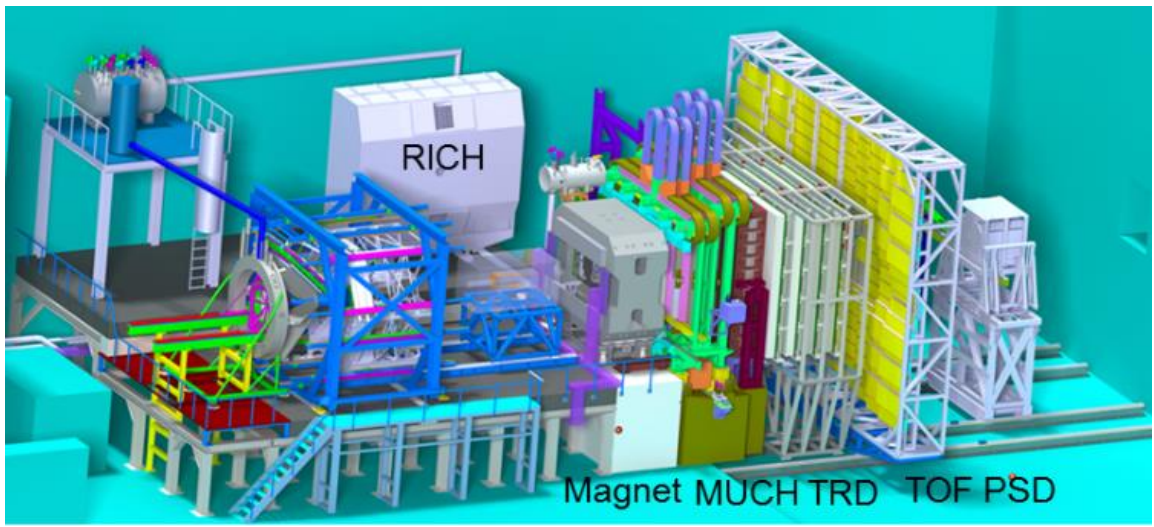

Fig. 11: Computer model of the HADES detector (left) and of the CBM setup (right). The beam enters from left. The detector systems will be operated alternatively.

In order to test the free-streaming data readout and acquisition system, a demonstrator experiment has been setup at SIS18 of GSI. This "miniCBM" comprises full-size prototype modules of each detector system including their read-out electronics, and the components of the DAQ system, including the optical links to theGeenIT cube. The DAQ system has been successfully operated with ion collisions at reaction rates up to 10 MHz. More information on the status of the CBM experiment including physics performances studies can be found [60].

4.3 The Nuclotron-based Ion Collider Facility (NICA)

The NICA complex is under construction on the premises of the Joint Institute for Nuclear Research (JINR) in Dubna, Russia [61]. The facility comprises a chain of three synchrotrons: the new Booster synchrotron, the existing Nuclotron with a magnetic rigidity 45 Tm, and the new collider ring, where collision energies of up to $\sqrt{s_{NN}}$ = 11 GeV and luminosities of $10^{27}$ cm$^{-2}$s$^{-1}$ can be reached for Au beams. The collider hosts two experimental setups: The Multiple Purpose Detector (MPD), where heavy-ion experiments will be performed, and the Spin Physics Detector (SPD), which is designed for experiments with beams of polarized protons and deuterons. The beam from the Nuclotron can also be sent directly into the experimental hall to the Baryonic Matter at Nuclotron (BM@N) detector setup, which is already in operation with light ion beams in order to study, for example, short-range correlations in inverse kinematics. The BM@N detector system is presently being upgraded for experiments on dense matter physics with Au ions of 3.8A GeV kinetic energy and intensities of up to about $10^9$ ions/s. A sketch of the NICA complex and a photo of the construction site from summer 2021 is shown in figure 12 and 13, respectively.

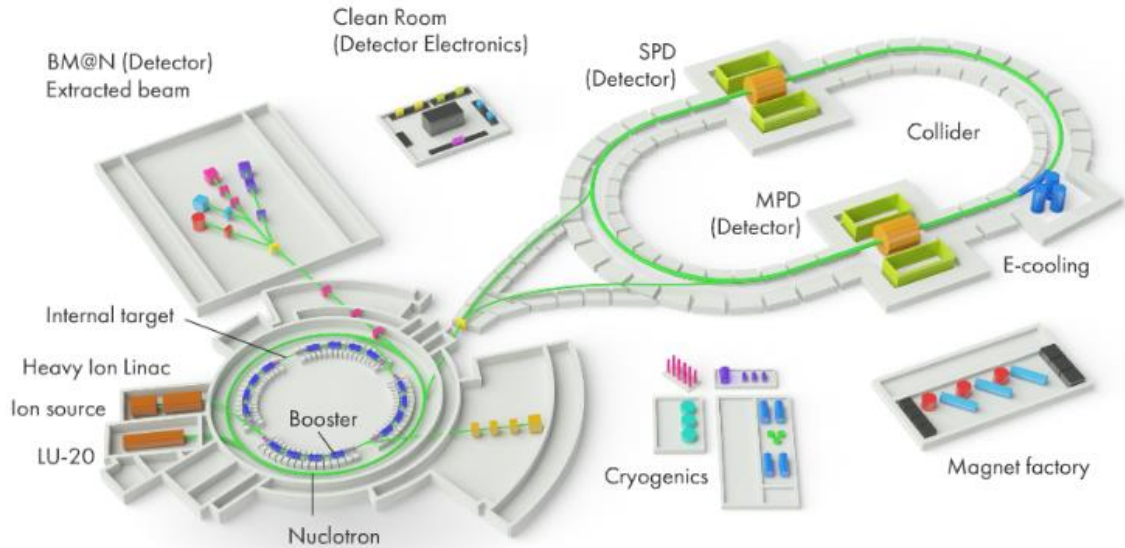

Fig.12: Sketch of the NICA complex [61].

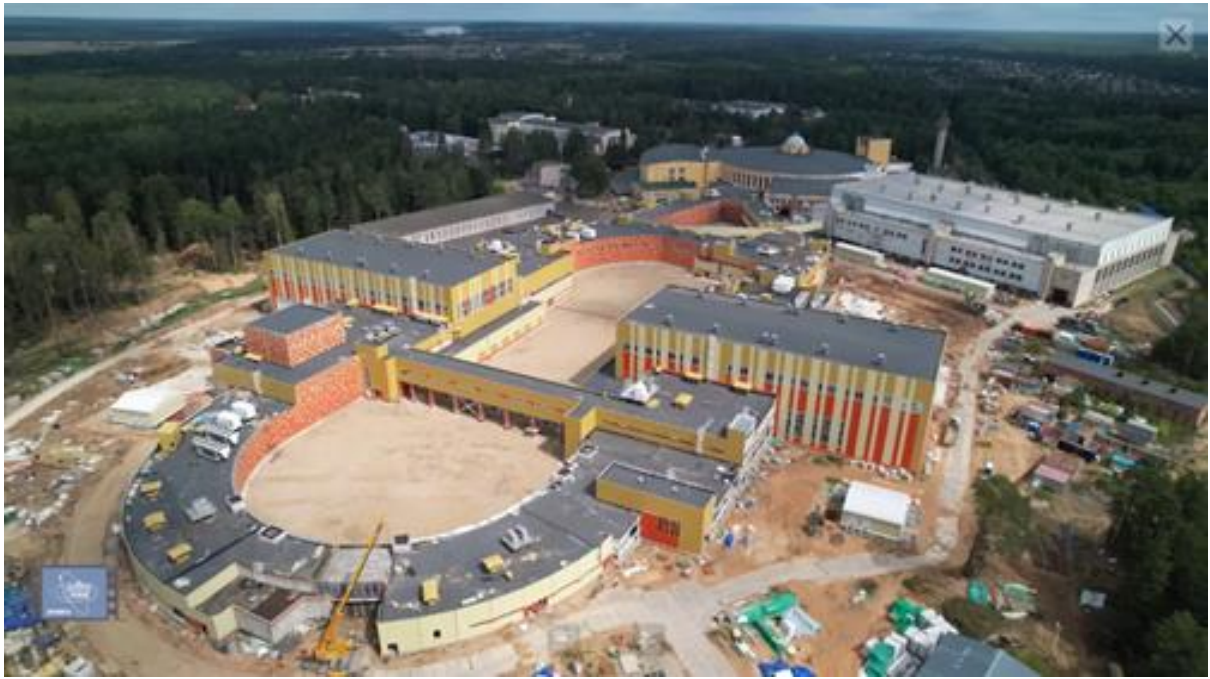

Fig. 13: Photo of the NICA construction site [61].

4.4 The MPD experiment at NICA

One of the two experimental setups in the NICA collider ring is the Multi-Purpose-Detector (MPD), which is devoted to the investigation of the QCD phase diagram [61,62]. The MPD experiment is designed to study light and heavy collision systems in the energy range from about $\sqrt{s_{NN}} = 4$ GeV (with low luminosity, see figure 8) up to $\sqrt{s_{NN}} = 11$ GeV with a luminosity of $10^{27}$ cm$^{-2}$s$^{-1}$, corresponding to a reaction rate of 6 kHz for Au+Au collisions. The MPD physics program includes the investigation of:

- event-by-event fluctuations in hadron production,

- femtoscopic correlations;
- collective flow of hadrons
- multi-strange hyperon production (yields, spectra)
- hypernuclei (yields, masses, lifetimes, spectra)
- photons and electron-positron pairs

As a typical collider experiment, the detector system comprises a Time-Projection Chamber (TPC) around the collision zone, surrounded by a Time-of-Flight (TOF) barrel and an Electromagnetic Calorimeter (ECAL). These detectors are installed in a 5 m long superconducting solenoid with an inner diameter of 4 m. As a second stage, the experiment will be upgraded by an Inner Tracking System (ITS) around the beam pipe, in order to identify the decay vertices of short-lived particles like mesons with open charm. The forward and backward hemispheres will be covered by Zero Degree Calorimeters (ZDC), Fast Forward Detectors (FFD), and Straw Tube Trackers (ECT). Figure 14 illustrates the layout of the MPD setup [61]. According to the present schedule, the collider and the MPD setup will be commissioned end of 2023.

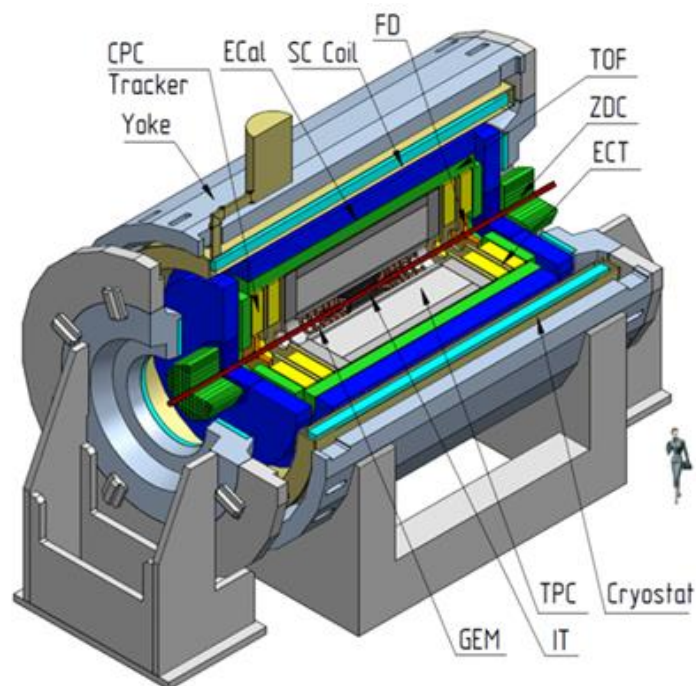

Fig. 14: Layout of the MPD experiment at NICA [61].

4.5. The BM@N experiment a NICA

The Baryonic Matter at Nuclotron (BM@N) fixed target experiment comprises a normal conducting dipole magnet with a gap of about 1 m, where the target station and the charged particle tracking detectors are installed. The first tracking system consists of three stations equipped with 42 two-coordinate silicon detectors, located between about 15 and 35 cm downstream the target. Downstream the silicon stations, six triple GEM stations are installed actually covering the upper hemisphere only. Two large drift chambers (DCH) provide tracking between the magnet and two TOF detector stations at 4m and 7m downstream the target, which consist of multi-Resistive Plate Chambers (mRPCs). Projectile spectator fragments are detected by a Zero-Degree Calorimeter, which provides information on the orientation of the reaction plane. This detector configuration is in operation since 2018 and has been used to measure lambda hyperons in C + C, Al and Cu collisions at a beam energy of 4A GeV [63]. A photo of this setup is shown in figure 15.

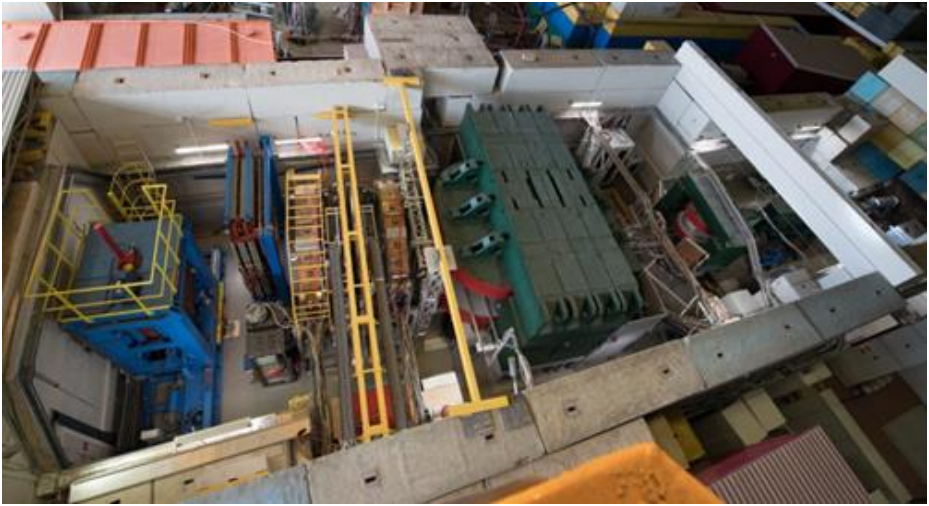

Fig. 15: Photo of the BM@N experimental setup. See text.

The BM@N setup has also been used for an experiment on short- range correlations (SRC). For this measurement, a liquid hydrogen target has been installed upstream the magnet, and two additional detector arms have been mounted to measure the knock-out protons under emission angles between 24° and 37°. Each arm consisted of two scintillator trigger counters (TC), a GEM station and a RPC wall. With this setup it was possible to perform a fully exclusive measurement in inverse kinematics with a 4A GeV C beam, where the residual A-2 nuclear system was probed by identification of all products of the reaction and $^{12}C + p \rightarrow 2p + ^{10}_{4}Be + p$ [64].

The BM@N setup is presently being upgraded for experiments with Au beams. This upgrade program comprises the installation of a highly granulated silicon tracking system, the completion of the GEM stations, the replacement of the drift chambers by highly granulated Cathode Strip Chambers (CSC), the exchange of the ZDC by a Forward Hadron Calorimeter (FHCal), and by the installation of beam pipes upstream and downstream the target. The development and the construction of the four stations of double-sided micro-strip silicon sensors is a joint development of the BM@N and the CBM collaboration, as the same detector type is used for the CBM experiment. The upgraded BM@N experiment will be able to contribute to the search for a phase transition at large baryon chemical potentials and to the study of the high density EOS by investigating light and heavy collision systems at beam energies up to 3.8A GeV for Au ions and up to 6A GeV for light ions [65]. The envisaged BM@N research program includes the following observables:

- multi-differential hadron distributions,
- event-by-event fluctuations and correlations,
- collective flow of identified particles,
- multi-strange hyperons,
- lambda hypernuclei

## 5. Summary and conclusion

The fundamental properties of strongly interacting matter at high baryon density, which govern the characteristics of compact stellar objects and the dynamics of supernova explosions and neutron star mergers, are still poorly known. In particular, the EOS of dense QCD matter, and its microscopic degrees-of-freedom are in the focus of present research activities, both in terrestrial laboratories and in space. The present constraint of the EOS up to densities of about 2 $\rho_0$ is provided by laboratory experiments with stable nuclei and heavy-ion beams. Pioneering heavy-ion experiments at intermediate beam energies, together with astronomical observations, have provided also the first information on the EOS at densities up to about 4 $\rho_0$. Future experiments with heavy-ion beams at FAIR and NICA are designed for precision measurements of observables, which are expected to be most sensitive to the high-density EOS, like the collective flow of hadrons, and subthreshold production of multi-strange (anti-) hyperons. Moreover, these experiments will shed light on the degrees-of-freedom of strongly interacting matter at neutron star core densities, and, hence, explore the yet almost uncharted region of the QCD phase diagram at large baryon chemical potentials. Both LQCD calculations and QCD based models restrict the location of a possible critical endpoint of a 1$^{st}$ order chiral transition from hadronic to quark-gluon matter to temperatures and baryon chemical potentials, which are accessible in heavy-ion collisions at FAIR-NICA energies. Here, the sensitive observables include excitation functions of particles containing strange and charm quarks, of higher order fluctuations of the proton number distribution, and of invariant-mass distributions of lepton pairs. The dilepton spectra will permit to extract the fireball temperature and the caloric curve of dense QCD matter, resulting in the discovery or the exclusion of a 1$^{st}$ order phase transition in dense QCD matter.

In conclusion, the future heavy-ion experiments at FAIR and NICA will not only complement astronomical observations on the high-density EOS, but will in addition explore the elementary structure of matter at neutron star core densities.

**Acknowledgement**

The author receives funding from the Europeans Union's Horizon 2020 research and innovation programme under grant agreement No. 871072.

**Postscript**

Due to the sanctions imposed on Russia because of the Russian attack on Ukraine, the close cooperation between FAIR and NICA is suspended, and the completion of the facilities might be delayed.

**References**

[1] Andronic A et al. 2018 Nature 561 321

[2] Bazavov A et al. (HotQCD Collaboration) 2019 Phys. Lett. B **795** 15

[3] Borsanyi S et al. Phys. Rev. Lett. 2020 **125** 052001

[4] Becattini F 2009, arXiv:0901.3643 [hep-ph];

[5] Cleymans J et al. 2006 Phys. Rev. C **73** 034905

[6] Stephanov M et al. 1998 Phys. Rev. Lett. 81 4816.

[7] Karsch F 2019 arXiv:1905.03936

[8] Ding H T et al. (HotQCD Collaboration) 2019 Phys. Rev. Lett. **123** 062002

9] Arsene, I et al. 2007 Phys. Rev. C **75** 24902

[10] Le Fevre A et al. 2016 Nucl. Phys. A **945** 112

[11] Russotto P et al. 2016 Phys. Rev. C **94** 034608

[12] Sturm C et al. 2001 Phys. Rev. Lett. **86** 39

[13] Fuchs C et al. 2001 Phys. Rev. Lett. **86** 1974

[14] Pinkenburg C et al. 1999 Phys. Rev. Lett. **83** 1295

[15] Danielewicz P, Lacey R and Lynch W G 2002 Science **298**

[16] Miller M C et al. ApJ. Lett. 2019 **887** L24

[17] LIGO and Virgo Collaborations, ApJ. Lett. 2017 848 L12.

[18] Huth S et al. arXiv:2107.06229

[19] Weise W 2019 JPS Conf. Proc. **26** 011002

[20] Fukushima K 2020 Phys. Rev. D **102** 096017

[21] Orsaria, M. et al., 2014 Phys. Rev. C **98** 015806.

[22] Gao F and Pawlowski J 2020 Phys. Rev. D **102** 034027

[23] Braun-Munzinger P et al. 2004 Phys. Lett. B**596** 6

[24] Andronic A et al. 2009 Acta Phys. Polon. B 40 1005

[25] Agakishiev G et al. 2011 Eur. Phys. J. A 47 21

[26] Matsui T and Satz H 1986 Phys. Lett. B**178** 416

[27] Abreu, M C et al. 2000 Phys. Lett. B **477** 28-26

[28] Braun-Munzinger P and Stachel J 2000 Phys. Lett. B**490** 196

[29] Andronic A et al. Phys. Lett. 2008 B**659** 149

[30] Adam J et al. ALICE Collaboration 2017 Phys. Lett. B**766** 212

[31] Linnyk O et al. 2007 Nucl. Phys. A **786** 183

[32] Steinheimer J et al. 2017 Phys. Rev. C **95** 014911

[33] Rapp R and van Hees H 2016 Phys. Lett. B**753** 586

[34] Specht H NA60 Collaboration 2010 AIP Conf. Proc. **1322** 160

[35] Adamczewski-Musch J et al. 2019 Nature Physics **15** 1040

[36] Hatta Y and Stephanov M. A. 2003 Phys. Rev. Lett. **91** 102003

[37] Adam J et al. STAR Collaboration 2021 Phys. Rev. Lett. **126** 092301

[38] Adamczewski-Musch J et al. HADES Collaboration 2020 Phys. Rev. C **102** 024914

[39] Abdallah M S et al. STAR collaboration 2021 arXiv:2112.00240

[40] Shuryak E and Torres-Rincon J M 2020 Eur. Phys. J. A **56** 241

[41] Youngblood D H et al., 1999 Phys. Rev. Lett. **82** 691

[42] Hartnack C and Aichelin J 2002 J. Phys. G **28** 1649

[43] Russotto P et al. 2016 Phys. Rev. C **94** 034608

[44] Li B-A 2017 Nuclear Physics News **27**, 7

[45] Liu H et al. 2000 Phys. Rev. Lett. **84** 5488

[46] Barrette J et al., 1997 Phys. Rev. C **56** 3254

[47] Cromartie H T et al. 2020 Nature Astronomy **4** 72

[48] Antoniadis J et al 2013 Science **340** 448

[49] Arzoumanian Z et al. 2018 ApJ **859** 47

[50] Li F et al. 2012 Phys. Rev. C **85** 064902

[51] Graef G et al. 2014, Phys. Rev. C **90** 064909  1592

[52] Aichelin J et al. 2020 Phys. Rev. C **101** 044905

[53] Estee J et al. 2021 Phys. Rev. Lett. **126** 162701